\documentclass[%
 reprint,
 superscriptaddress,
 amsmath,amssymb,
 aps,
]{revtex4-2}

\usepackage{url}
\usepackage{graphicx}
\usepackage{dcolumn}
\usepackage{bm}
\usepackage{hyperref}
\usepackage[autostyle]{csquotes}

\usepackage{physics}
\usepackage{mathtools}
\usepackage{comment}
\begin{document}






\title{Spin-Phonon-Photon Strong Coupling in a Piezoelectric Nanocavity}

\author{Hamza Raniwala}
\affiliation{Department of Electrical Engineering and Computer Science, Massachusetts Institute of Technology, Cambridge, MA 02139, USA}

\author{Stefan Krastanov}
\affiliation{Department of Electrical Engineering and Computer Science, Massachusetts Institute of Technology, Cambridge, MA 02139, USA}

\author{Lisa Hackett}
\affiliation{Sandia National Laboratories, Albuqurque, NM, USA}

\author{Matt Eichenfield}
\affiliation{Sandia National Laboratories, Albuqurque, NM, USA}

\author{Dirk R. Englund}%
\affiliation{Department of Electrical Engineering and Computer Science, Massachusetts Institute of Technology, Cambridge, MA 02139, USA}\affiliation{Research Laboratory of Electronics, Massachusetts Institute of Technology, Cambridge, MA 02139, USA}

\author{Matthew E. Trusheim}
\affiliation{Department of Electrical Engineering and Computer Science, Massachusetts Institute of Technology, Cambridge, MA 02139, USA}
\affiliation{U.S. Army Research Laboratory, Sensors and Electron Devices Directorate, Adelphi, Maryland 20783, USA} 

\date{\today}

\begin{abstract}
We introduce a hybrid tripartite quantum system for strong coupling between a semiconductor spin, a mechanical phonon, and a microwave photon. Consisting of a piezoelectric resonator with an integrated diamond strain concentrator, this system achieves microwave-acoustic and spin-acoustic coupling rates $\sim$MHz or greater, allowing for simultaneous ultra-high cooperativities ($\sim 10^3$ and $\sim 10^2$, respectively). From finite-element modeling and master equation simulations, we estimate photon-to-spin quantum state transfer fidelities exceeding 0.97 based on separately demonstrated device parameters. We anticipate that this device will enable hybrid quantum architectures that leverage the advantages of both superconducting circuits and solid-state spins for information processing, memory, and networking.




\end{abstract}

\maketitle



Solid-state quantum systems based on superconductors and spins are leading platforms that offer complementary advantages in quantum computing and networking. Superconducting quantum processors enable fast and high-fidelity entangling gates \cite{jurcevic2021demonstration,arute2020hartree}, but challenges remain in quantum memory time and long-distance networking. Conversely, atom-like emitters in solid-state have demonstrated long spin coherence time, efficient spin-photon interfaces for long-distance entanglement, and high readout fidelity \cite{steiner2010universal,lachlan2014all,trusheim2020transform,bhaskar2020experimental, pla2013high, nagy2019high}. Coupling these modalities is therefore an exciting direction in quantum information science. 

Previous studies using magnetic coupling between microwave (MW) photons and spins have been limited to multi-spin ensemble interactions \cite{ranjan2013probing, xiang2013hybrid, kubo2011hybrid, zhu2011coherent, sigillito2014fast, grezes2016towards, dold2019high}  due to low spin-magnetic susceptibility and the low magnetic energy density of MW resonators \cite{carter2015spin, rabl2009strong, angerer2018superradiant}. Alternate experiments and proposals rely on coupling via intermediate acoustic modes \cite{schuetz2015universal, neuman2020phononic,maity2020coherent}, which have experimentally demonstrated large coupling to superconducting circuits \cite{o2010quantum, arrangoiz2016engineering, arrangoiz2018coupling, arrangoiz2019strong, peterson2019ultrastrong} and are predicted to have large coupling to diamond quantum emitters \cite{kuzyk2018scaling, li2019honeycomblike, wang2020coupling, rabl2010quantum, lekavicius2019diamond, joe2021diamond,raniwala2022spin}. However, designing a device that strongly couples one phonon to both one MW photon and to one spin -- enabling an efficient MW photon-to-spin interface -- remains an outstanding challenge.



Here we address this problem through the co-design of a scandium-doped aluminum nitride (ScAlN) Lamb wave resonator with a heterogeneously-integrated diamond thin film. This structure piezoelectrically couples a MW photon and acoustic phonon while concentrating strain at the location of a diamond quantum emitter. Through finite-element modeling, we predict photon-phonon coupling $\sim10$ MHz concurrent with phonon-spin coupling $\sim3$ MHz. These rates yield photon-phonon and phonon-spin cooperativities of order $10^4$ assuming demonstrated lifetimes of spins, mechanical resonators, and superconducting circuits \cite{kjaergaard2020superconducting,devoret2013superconducting}. We explore state transfer protocols via quantum master equation (QME) simulations and show that this device can achieve photon-to-spin transduction fidelity $F > 0.97$ with conservative hardware parameters. We find that performance of these schemes is likely limited by two-level system (TLS) loss in current piezoelectrics. An improvement in piezoelectric TLS loss rates to that of silicon will pave the way towards SC-spin state transduction with $F > 0.99$.

We consider a coupled tripartite system consisting of a superconducting circuit (SC), acoustic phonon, and Group-IV electron spin with the Hamiltonian (see \cite{Supporting} for detailed derivation)

\begin{align}
    \begin{multlined}
    \frac{\hat{H}}{\hbar} = \frac{\omega_{sc}}{2}\hat{\sigma}^{z}_{sc}  + \omega_p \hat{a}_p^\dag \hat{a}_p + \frac{\omega_{e}}{2}\hat{\sigma}^{z}_{e} \\
    + g_{sc,p}\left( \hat{\sigma}^{+}_{sc} \hat{a}_p + \hat{\sigma}^{-}_{sc}\hat{a}_p^\dag\right) + g_{p,e}\left(\hat{\sigma}^{+}_{e}\hat{a}_p + \hat{\sigma}^{-}_{e}\hat{a}_p^\dag\right).
    \end{multlined}
    \label{system Hamiltonian}
\end{align}

\begin{figure}
\centering
	\includegraphics[width=\linewidth]{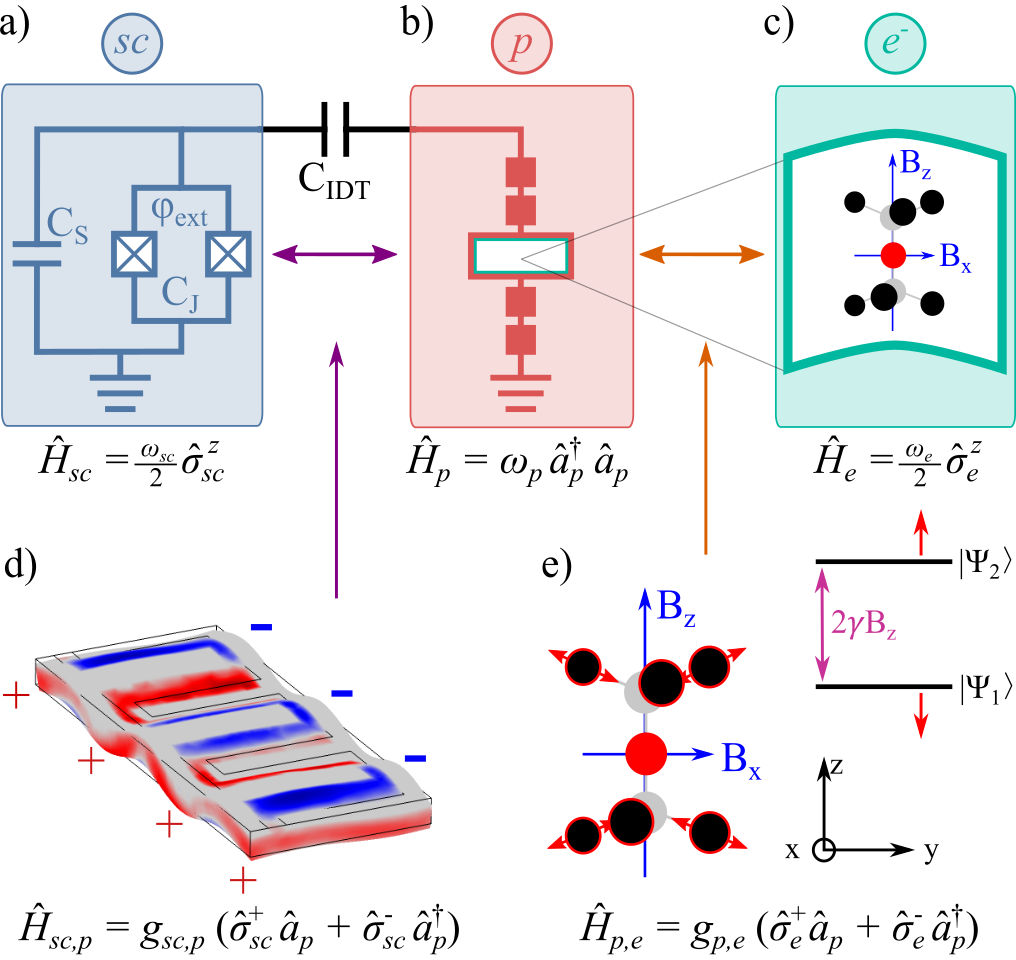}
	\caption{Coupled SC-phonon-spin quantum system. (a-c) depict the uncoupled modes of the (a) superconducting qubit with Josephson capacitance $C_J$, shunt capacitance $C_S$, and external flux bias $\phi_{ext}$; (b) acoustic mode capacitively coupled by $C_{IDT}$; and (c) diamond quantum emitter. (d) Piezoelectric interaction, where the color indicates the electric field profile under mechanical displacement. 
	(e) Spin-strain coupling resulting from modulating the inter-atomic distance of the quantum emitter via mechanical strain under an external $B$ field $\textbf{B} = B_x\hat{\textbf{x}} + B_z\hat{\textbf{z}}$ with spin-gyromagnetic ratio $\gamma$.
	}
	\label{fig 1: schematic}
\end{figure}

Here, the SC frequency $\omega_{sc}$ is defined by the transmon Josephson and shunt capacitances, the spin frequency $\omega_{e}$ is given by the Zeeman splitting of the electron spin states, and the acoustic frequency $\omega_p$ is defined by the acoustic resonator geometry. The first three terms of this equation describe the energies of the uncoupled modes of the devices (Fig.~\ref{fig 1: schematic}(a-c)) while the fourth and fifth terms describe the interaction dynamics. Generally, SCs feature $\omega_{sc} \sim 4-6$ GHz \cite{krantz2019quantum}. Electron spin resonant frequencies can be arbitrarily set by an external magnetic field; to match this frequency range, fields $\sim0.1$ T are required \cite{hepp2014electronic}. The coupling coefficient $g_{sc,p}$ is physically governed by the piezoelectric effect, whereby a strain field produces an electric response and vice versa (Fig.~\ref{fig 1: schematic}(d)). This interaction is described by the strain-charge equations
\begin{align}
    S_{ij} &= s_{ijkl}T_{kl} + d_{kij}E_{k}, \\
    D_{i} &= d_{ijk}T_{ij} + \epsilon_{ik}E_{k},
    \label{strain charge relations}
\end{align}
where $s_{ijkl}$ and $d_{ijk}$ are the elastic and piezoelectric coefficient tensors of the resonator's piezoelectric material, $S_{ij}$ and $T_{ij}$ are the stress and strain fields, and $E_{i}$ and $D_{i}$ are the electric and displacement fields. At single quantum levels, a MW photon in the SC will generate an electric field in the volume of the piezoelectric resonator described by
\begin{equation}
    \textbf{e}_{sc}(\textbf{r}) = \sqrt{\left(\frac{\hbar \omega_{sc}}{\left(C_S + C_J + C_{IDT}\right) V_{app}^2/2}\right)}\textbf{E}_{IDT}(\textbf{r})e^{-i\omega_{sc} t},
\end{equation}
where $\textbf{E}_{IDT}(\textbf{r})$ is the electric field profile of the IDT for an arbitrary applied voltage $V_{app}$, and the capacitances are indicated in Fig.~\ref{fig 1: schematic}. Since $C_S$ is typically much larger than $C_{IDT}$ and $C_J$ for transmon qubit configurations, the MW photon energy is largely contained in the shunt capacitor. Similarly, a phonon in the piezoelectric resonator will produce a strain field described by
\begin{equation}
    \textbf{t}_{p}(\textbf{r}) = \sqrt{\left(\frac{\hbar\omega_{p}}{\int_V dV\;\textbf{s}(\textbf{r})\abs{\textbf{T}_p(\textbf{r})}^2/2}\right)}\textbf{T}_{p}(\textbf{r})e^{-i\omega_{p} t},
\end{equation}

where $\textbf{T}_p(\textbf{r})$ is the strain profile of the acoustic mode for an arbitrary mechanical displacement. Following \eqref{strain charge relations}, $\textbf{t}_p(\textbf{r})$ will produce an electric displacement field given by $\textbf{d}\cdot\textbf{t}_p(\textbf{r})$, where $\textbf{d}$ is the piezoelectric coefficient tensor. Then the coupling $g_{sc,p}$ will be determined by the overlap integral between $\textbf{e}_{sc}(\textbf{r})$ and $\textbf{d}\cdot \textbf{t}_p(\textbf{r})$ \cite{zou2016cavity},
\begin{equation}
    g_{sc,p} = \frac{1}{2\hbar}\int_{V} dV \left(\textbf{t}^*_{p}(\textbf{r})\cdot\textbf{d}^T\cdot\textbf{e}_{sc}(\textbf{r}) + \textbf{e}^*_{sc}(\textbf{r})\cdot \textbf{d}\cdot\textbf{t}_p(\textbf{r})\right).
    \label{gscp formula}
\end{equation}

\begin{figure*}[htb!]
\centering
	\includegraphics[width=\linewidth]{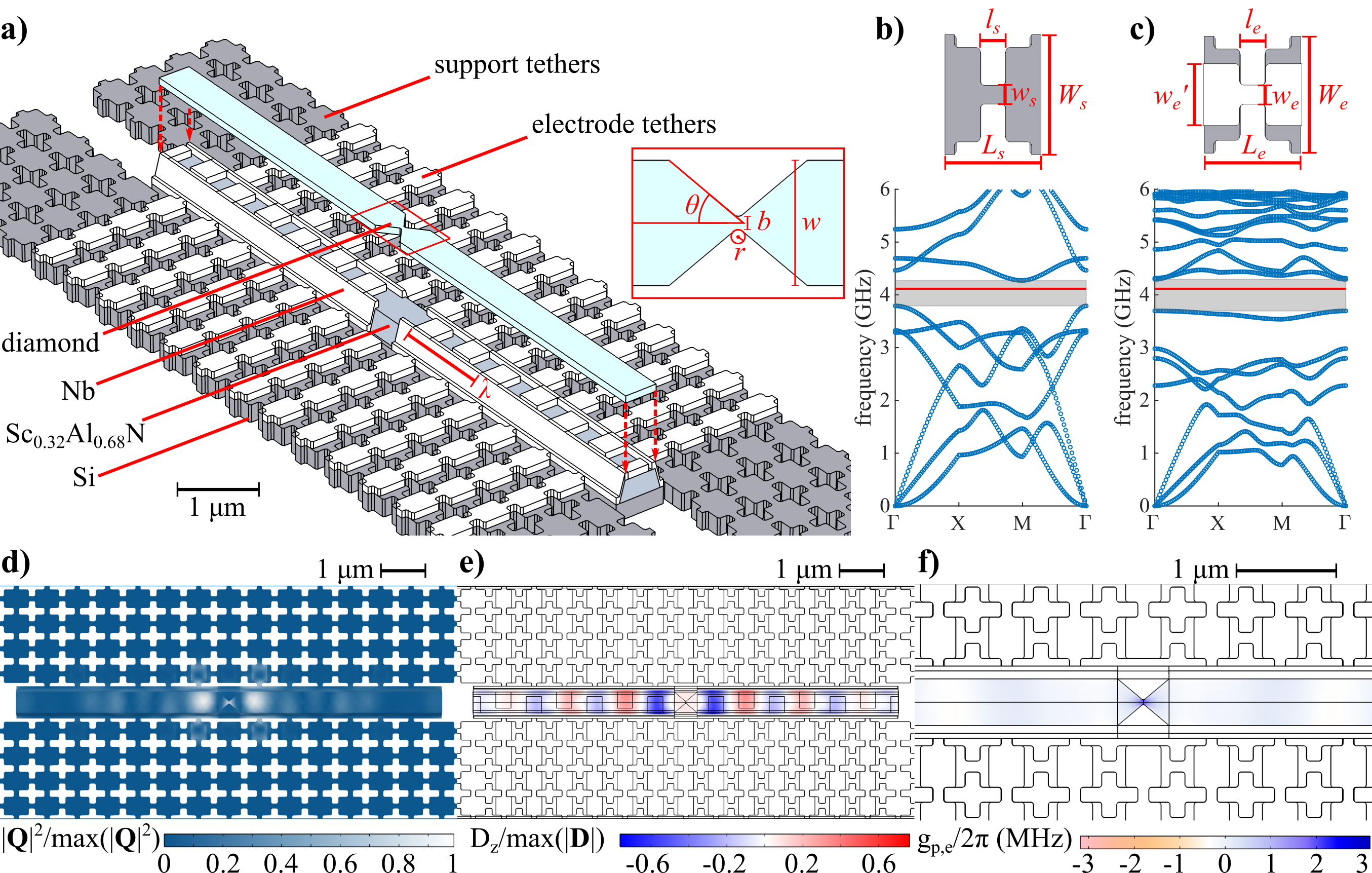}
	\caption{Electromechanical transducer design. (a) Lamb wave resonator and relevant design parameters. In this Letter, the resonator geometry is parametrized by $(\lambda, w, t_d,t_{Al}, t_{AlN},t_{Si}) = (1370, 465, 100, 100, 300, 250)$[nm] ($t_i$ is the thickness of layer $i$), with the diamond taper defined by $(b, r,\theta) = (40\text{ nm}, 25\text{ nm}, 50^\circ)$. The phononic tethers are of two types: support tethers defined by $(W_s, L_s, w_s, l_s) = (705, 565, 110 150)$[nm] and electrode tethers defined by $(W_e, L_e, w_e, l_e) = (685, 565, 110, 150)$[nm]. 
	(b,c) Phononic band structure of the support (c: electrode) tethers, with a 500 MHz band gap indicated in grey shading and the resonant frequency of the device indicated with the red line. (d) Normalized mechanical displacement of the resonator. (e) Induced piezoelectric displacement field at the central slice of the piezoelectric layer. (f) Spatial profile of $g_{p,e}$ at the center slice of the diamond layer, assuming a magnetic field of 0.18 T. 
	}
	\label{fig 2: diagram}
\end{figure*}

The spin-phonon coupling $g_{p,e}$ results from the spin-strain susceptibility $\mathbf\chi_{spin}$ of quantum emitters in a strain field \cite{hepp2014electronic,meesala2016enhanced,meesala2018strain}. For a single-phonon strain profile $\textbf{t}_{p}$, the resulting coupling is $g_{spin}(\textbf{r}) = \mathbf{\chi}_{spin}\cdot\textbf{t}_p(\textbf{r})$. In Group IV emitters in diamond, $\chi_{spin}$ depends heavily on the spin-orbit mixing enabled by an off-axis magnetic field (see \cite{Supporting}) and primarily interacts with transverse strain in the emitter frame \cite{hepp2014electronic}. Therefore, for the rest of this analysis, we set this expression to be 
\begin{equation}
    g_{p,e}(\textbf{r}) = \chi_{eff}(t'_{xx}(\textbf{r}) - t'_{yy}(\textbf{r})),
    \label{gpe formula}
\end{equation} 
where $\textbf{t}'(\textbf{r})$ is the single-phonon strain profile in the coordinate system of the emitter and $\chi_{eff} \approx 0.28$ PHz/strain \cite{meesala2018strain}. 
To implement the device in Fig.\ref{fig 1: schematic}, we require a platform with (i) superconductivity, (ii) piezoelectricity, (iii) acoustic cavities, and (iv) strain transfer to diamond emitters. To address (i-ii), we propose a silicon-on-insulator (SOI) platform with a thin-film deposition of scandium-doped aluminum nitride (ScAlN). This material system allows for superconducting qubits and piezoelectrics to co-inhabit one chip \cite{keller2017transmon,keller2017superconducting}. To answer (iii-iv), we co-design a Nb-on-Sc$_{0.32}$Al$_{0.68}$N-on-SOI piezoelectric resonator with a heterogeneously integrated diamond thin membrane. We propose Niobium (Nb) as a well-characterized superconductor with high $H_{c1} = 0.18$ T and $H_{c2} = 2$ T \cite{saito2001critical, kerchner1981critical, finnemore1966superconducting}, as required for operation with the spin. SOI platforms have previously been used for piezoelectric resonators \cite{lobl2001materials,loebl2003piezoelectric}, and diamond-AlN interfaces have been used to acoustically drive emitters in diamond \cite{golter2014optically,golter2016coupling,golter2016optomechanical}. ScAlN further boosts the piezoelectric coefficient of AlN, allowing us to achieve a stronger interaction \cite{akiyama2009influence,kurz2019experimental}. 

We present the resonator design in Fig.~\ref{fig 2: diagram}. Our device is based on Lamb wave resonators, which produce standing acoustic waves dependent on electrode periodicity $\lambda$ and material thickness \cite{bjurstrom2005lateral, lin2010temperature, konno2013scaln}. We localize the strain in the diamond thin film using a central ``defect cell" (Fig. \ref{fig 2: diagram}a inset) with a suspended taper. To maintain high quality factors, we tether the Lamb wave resonator via phononic crystal tethers placed at displacement nodes of the box. \cite{mirhosseini2020superconducting}. We further propose an angled ScAlN sidewall in the transducer (15$^\circ$ from normal) that allows the electrodes to "climb" on top of the ScAlN film, rather than requiring a continuous piezoelectric layer over the phononic tethers. This both facilitates the design of wide-bandgap phononic tethers and is compatible with current fabrication techniques.

 To calculate device performance, we simulate the architecture using the finite element method (FEM) in COMSOL to produce the phononic tether band structures and mode profiles (Fig. \ref{fig 2: diagram}b-e). The tether band structure exhibits a 500 MHz bandgap around the device's $\approx$ 4.11 GHz resonant mode. This frequency is desirable as it falls near the central operating range of most superconducting qubits \cite{krantz2019quantum}. Additionally, the 4.11 GHz resonant mode is itself isolated from other acoustic modes of the system by $\sim$56 MHz, which is enough to neglect parastic couplings and treat the transducer in the single-mode approximation (see SI). Fig.~\ref{fig 2: diagram}(d-e) show the mechanical and electrical displacement fields of this mode, from which we derive $\mathbf{e}_{sc}(\mathbf{r})$ and $\mathbf{t}_p(\mathbf{r})$, respectively. 
We calculate a $g_{sc,p}\approx7.0-20.5$ MHz (for a shunt capacitance of 65-190 fF, corresponding to $100\text{ MHz}<E_{C}/h<300\text{ MHz}$\cite{krantz2019quantum}) and a maximum $g_{p,e}\approx3.2$ MHz according to Equations \eqref{gscp formula} and \eqref{gpe formula}. The strain maximum occurs at the edges of the central diamond taper, which maximizes $g_{p,e}$ (Fig.~\ref{fig 2: diagram}f). 
\begin{figure*}[htb!]
\centering
	\includegraphics[width=\linewidth]{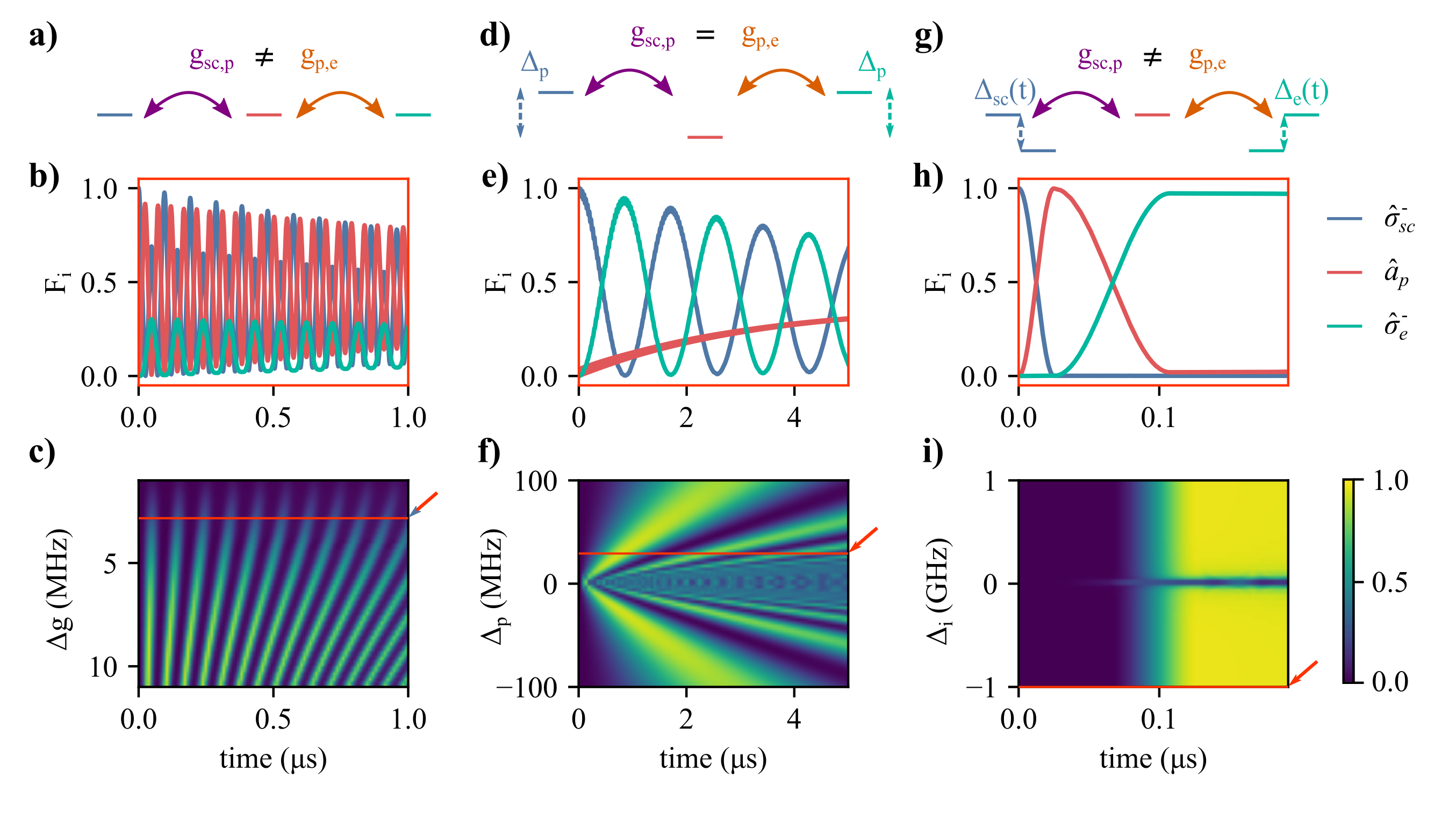}
	\caption{Analysis of the coupled SC-phonon-spin system under different protocols: (a-c) uncontrolled time evolution, when all modes are on resonance and coupling rates are maximized; (d-f) time evolution detuned from the acoustic resonance, which allows for state transfer through virtual phonon excitation; and (g-h) time evolution under detuning control, which allows for controlled Rabi flops across the modes. Plots (b,e,h) depict the population dynamics of each mode for the above protocols. Plots (c,f,i) show the spin population over time for the variable parameter of the procedure, with operational points for plots (b,e,h) indicated with orange lines. (c) shows population for a given $\Delta g$, (f) shows population for achievable phonon detuning $\Delta_p$, and (i) shows performance for unused mode detuning $\Delta_i$ during each Rabi swap.}
	\label{fig 3: protocol}
\end{figure*}

In Fig.~\ref{fig 3: protocol}, we explore different protocols for quantum transduction from an initialized SC to a spin. The time evolution of the system when initialized in the $\rho_0 = \ket{100}\bra{100}$ state (where the indices consecutively refer to the state of the SC, the Fock state of the phonon, and the z-projection of the spin) is calculated using the Lindblad master equation,
\begin{equation}
\begin{multlined}
    \frac{d}{dt}\rho = -\frac{i}{\hbar}[\rho,\hat{H}(t)] + \kappa_{p}\left(\hat{a}_p^\dag\rho\hat{a}_p - \hat{a}_p^\dag \hat{a}_p \rho - \rho\hat{a}_p^\dag\hat{a}_p\right)\\+
    \kappa_{sc}\left(\hat{\sigma}_{sc}^{-}\rho\hat{\sigma}_{sc}^{+} - \hat{\sigma}_{sc}^{+}\hat{\sigma}_{sc}^{-}\rho - \rho\hat{\sigma}_{sc}^{+}\hat{\sigma}_{sc}^{-}\right)\\+
    \kappa_{e}\left(\hat{\sigma}_{e}^{-}\rho\hat{\sigma}_{e}^{+} - \hat{\sigma}_{e}^{+}\hat{\sigma}_{e}^{-}\rho - \rho\hat{\sigma}_{e}^{+}\hat{\sigma}_{e}^{-}\right).
    \end{multlined}
\end{equation}
where the Hamiltonian in a frame rotating at rate $\omega_p$ is
\begin{align}
    \begin{multlined}
    \frac{\hat{H}(t)}{\hbar} = \frac{\Delta_{sc}(t)}{2}\hat{\sigma}^{z}_{sc}+  \frac{\Delta_{e}(t)}{2}\hat{\sigma}^{z}_{e} \\
    + g_{sc,p}\left( \hat{\sigma}^{+}_{sc} \hat{a}_p + \hat{\sigma}^{-}_{sc}\hat{a}_p^\dag\right) + g_{p,e}\left(\hat{\sigma}^{+}_{e}\hat{a}_p + \hat{\sigma}^{-}_{e}\hat{a}_p^\dag\right).
    \end{multlined}
    \label{evolving Hamiltonian}
\end{align}
Here, $\Delta_{sc}(t)\equiv \omega_{sc}(t) - \omega_p$ is the superconducting qubit detuning and $\Delta_{e}(t)\equiv \omega_{e}(t) - \omega_p$ is the spin detuning at time $t$. The use of time-varying detuning can be easily implemented, e.g. via on-chip flux bias lines \cite{sillanpaa2007coherent,strand2013first, mckay2016universal}, unlike time-varying coupling rates explored in previous works \cite{neuman2020phononic}. We account for dephasing in each mode with conservative estimates on decoherence rates $\frac{\kappa_{sc}}{2\pi} = 100$ kHz, $\frac{\kappa_p}{2\pi} = \frac{\omega_p}{2\pi Q} \approx 40$ kHz, and $\frac{\kappa_e}{2\pi} = 1$ MHz \cite{devoret2013superconducting, kjaergaard2020superconducting,  premkumar2021microscopic, sukachev2017silicon, pingault2017coherent}. As cryogenic operation of ScAlN-on-SOI acoustic resonators--as well as diamond hybrid intergration on said devices--has not been previously explored, we further discuss prospects for $Q_{mech}$ below.

Fig.~\ref{fig 3: protocol}(b,e,h) plot the state transfer fidelity $F_i\equiv \bra{\psi_{i}}\rho(t)\ket{\psi_{i}}$ to the target state $\ket{\psi_{i}} = \ket{1_i}$ under different conditions. In Fig.~\ref{fig 3: protocol}a where the modes are all resonant ($\omega_{sc}=\omega_{p}=\omega_{e} = 4.11$ GHz), and $g_{sc,p}/2\pi=10$ MHz, $F_{e}$ is poor due to the mismatch $\Delta g(g_{p,e}) = g_{sc,p}- g_{p,e}$ (Fig.~\ref{fig 3: protocol}c). Assuming one reduces $g_{sc,p}$ or $g_{p,e}$, for example by increase the qubit shunt capacitance $C_S$ or reducing the transverse magnetic field, $F_e$ may increase at the cost of maximum coupling rates. 


In Fig.~\ref{fig 3: protocol}b we detune the phonon mode by $\Delta_p \equiv \omega_{p} - \omega_{sc}$ where $\omega_{sc} = \omega_{e}$ and keep the coupling rates matched at 3.0 MHz. In this case, $F_e \sim 0.95$ via virtual excitation of the phonon mode, if the phonon mode is detuned by $30$ MHz. This protocol generates very low population in the phonon mode, primarily exchanging states between the superconducting qubit and spin. If the phonon mode is lossy, this transduction method is then preferred. However, while this protocol features wider efficiency peaks in time, which may require less stringent pulse control (see Fig.~\ref{fig 3: protocol}e), it does not overcome the issue of coupling imbalance and additionally suffers from decoherence of the superconducting qubit and spin modes over a longer protocol time  (Fig.~\ref{fig 3: protocol}f).

Fig.~\ref{fig 3: protocol}g shows the optimal solution, assuming control over $\Delta_{sc}(t)$ and $\Delta_{e}(t)$, in a double Rabi-flop protocol. During this protocol, it is assumed that $g_{sc,p}/2\pi=10$ MHz (which overcomes losses during the Rabi flop while still allowing mode isolation during the next flop) and $g_{p,e}/2\pi = 3.0$ MHz. We also assume $\Delta_{sc}(t) = 0$ and $0 \text{ MHz}\leq \Delta_{e}(t) \leq 1$ GHz for $t \in \{0, \pi/(2 g_{sc,p})\}$--the duration of a Rabi flop between the SC and phonon. Then, $\Delta_{sc}(t) = \Delta_{e}(t=0)$ MHz and $\Delta_{e}(t) = 0$ for $t \in \{\pi/(2 g_{sc,p}), \pi/(2 g_{sc,p}) + \pi/(2 g_{p,e})\}$--the duration of a Rabi flop between the phonon and spin. This sequentially transfers states between the modes (Fig.~\ref{fig 3: protocol}h), and for $\Delta_{e}(t=0) > 500$ MHz, can achieve $F_e > 0.97$ (Fig.~\ref{fig 3: protocol}i; for $\Delta_{i} = 1.0$ GHz, $F_e = 0.971$). In this protocol, we have neglected the losses that can occur when varying $\Delta_{sc}$ and $\Delta_{e}$. In reality, one has to select a pair of $\Delta_{sc}$ and $\Delta_{e}$ that do not fall on resonance with another acoustic mode of the system to prevent Rabi oscillations between the SC qubit or electron spin and an undesired acoustic mode (see SI for more details).

Each of these scenarios achieves transduction to the spin with high fidelity. The third scenario allows the quantum state to persist in the spin without continued interaction with the acoustic or SC modes. While in this state, the electron spin can access other degrees of freedom (e.g. $^{13}C$ spins \cite{metsch2019initialization, maity2021coherent}).

Since acoustic losses and therefore the total mechanical quality factor $Q_{mech}$ are difficult to predict from first principles, we evaluate the transduction fidelity $F_{e}$ of each protocol in different regimes of $Q_{mech}$ in Fig. 4. Here, protocol 1 is the resonant protocol with $g_{sc,p}=g_{p,e}=3$ MHz; protocol 2 is the virtual excitation protocol with identical $g_{sc,p}$ at a detuning of $30$ MHz; and protocol 3 is the Rabi protocol with $g_{sc,p}=10$ MHz, $g_{p,e}=3$ MHz, and $\Delta_{e}(t=0)=1$ GHz. $Q_{mech}$ is the inverse sum of three components,
\begin{equation}
    Q_{mech} = \left(Q_{c}^{-1} + \sum_{i} p_i \left(Q_{TLS, i}\right)^{-1} + Q_{A}^{-1}\right)^{-1}.
\end{equation}
Here, $Q_{c}$ is the mechanical clamping loss, which we can engineer to be non-limiting (see \cite{Supporting}), and $Q_{A}$ is the Akhieser loss-related $Q$, which at millikelvin temperatures is negligible \cite{chan2012laser}. These two losses are well-described for analogous systems; in contrast, $Q_{TLS, i}$--the dielectric loss-related $Q$--is harder to predict. These $Qs$ depends on the number of quasi-particles or TLSs trapped in each of the device's material interfaces and are weighted by the electric field participation $p_i$ in each interface. Given this uncertainty in $Q_{TLS_i}$, we lay out the protocol hierarchy as a function of the overall $Q_{mech}$:
\begin{itemize}
    \item If $Q_{mech} \lesssim 2\times10^3$, protocol 2 is superior.
    \item If $2\times10^3 \lesssim Q_{mech} \lesssim 5\times10^5$, protocol 1 is superior.
    \item If $5\times 10^5 \lesssim Q_{mech}$, protocol 3 is superior.
\end{itemize} In existing hardware, the largest challenge to reach the high-fidelity regime ($F \gtrsim 0.99$) is reducing dielectric loss in the thin-film piezoelectric, as indicated by published intrinsic quality factors of, e.g., monolithic aluminum nitride or lithium niobate resonators \cite{fan2013aluminum, wollack2021loss}. So, while current hardware may encourage us to utilize the virtual coupling protocol for coupling through a lossy intermediary phononic mode, future iterations of this scheme with improved materials and interfaces can expect to break the 0.99 transduction fidelity barrier using a resonant protocol. At this fidelity, SC-spin transduction would surpass the 1\% error correction thresholds of common codes and thus be compatible with scalable quantum information processing schemes \cite{kitaev2003fault,raussendorf2007topological,wang2009threshold}. 

An open question remains in the bonding strength between the diamond thin film and underlying resonator, which, if poor, can incur additional losses. However, for single-phonon occupation, the Van der Waals static frictional force exceeds the strain-generated force on the resonator. 

\begin{figure}
    \centering
    \includegraphics[width=\linewidth]{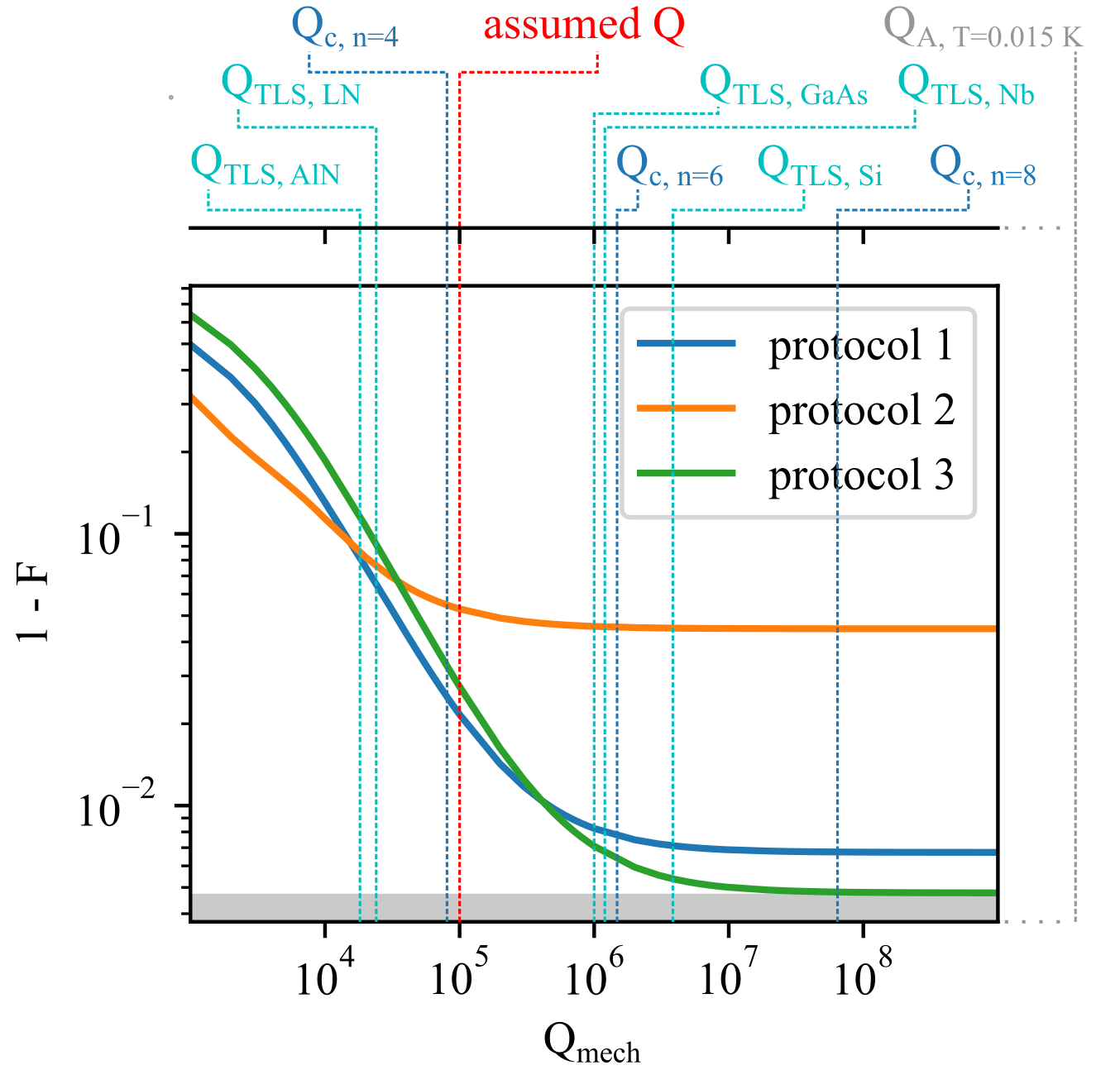}
    \caption{Sweep of protocol performance as a function of the total quality factor of the mechanical resonator. TLS-limited $Q$s for Si \cite{melville2020comparison, woods2019determining}, AlN \cite{fan2013aluminum}, Nb \cite{mcrae2020materials}, and alternatives in GaAs \cite{mcrae2020materials} and LN \cite{wollack2021loss} are in cyan. This device's clamping-limited $Q$s as a function of tether number are listed in blue. Akhiezer losses (grey) are non-dominant at $T = 0.015$ K. Finally, our assumed $Q \approx 10^5$ is in red. The $F > 0.995$ regime (dark grey) requires better SCs and spins to achieve.}
    \label{fig 4: Q factors}
\end{figure}


Ultimately, we have proposed a resonator architecture capable of simultaneously coupling a microwave photonic mode from a superconducting circuit and an electronic spin from a solid state color center to a single phonon. 
For our calculated coupling parameters and conservatively assumed $Q$s across the three modes, we expect SC-phonon cooperativity $C_{sc,p} = \frac{4 g_{sc,p}^2}{\kappa_{sc}\kappa_{p}} \sim 4\times 10^3$ and similarly, spin-phonon cooperativity $C_{p,e} = \frac{4 g_{p,e}^2}{\kappa_{p}\kappa_{e}} \sim 10^2$.
This doubly strongly-coupled architecture has a number of uses. Firstly, it can provide superconducting circuit qubits access to a long-lived quantum memory in the form of a nuclear spin register surrounding the electron spin. Secondly, this resonator can grant superconducting circuit qubits a spin-photon interface for efficient coupling to fiber optical quantum networks. 
Finally, by multiplexing each SC with several acoustic resonators and each acoustic resonator with several spins, this architecture can yield a memory bank of quantum memories for computational superconducting circuits (see \cite{Supporting} for a more detailed discussion).
We believe introducing this quantum transducer into existing superconducting circuits is a large step towards developing a specialized hybrid quantum computer with fast superconducting qubits for processing and slow, long-lived memory qubits in the solid state for storage and communication.





\begin{acknowledgements}
The authors would like to thank Ian Christen for insightful comments and discussions pertaining to this research. HR acknowledges funding from the NDSEG fellowship. HR, SK, ME, and DRE acknowledge funding from the MITRE Corporation and the NSF Center for Ultracold Atoms. MET acknowledges support from the Army Research Laboratory ENIAC Distinguished Postdoctoral Fellowship. ME and LH acknowledge support from Sandia National Laboratories.
\end{acknowledgements}

\bibliography{references}

\clearpage
\onecolumngrid 
\appendix
\section{Theoretical Analysis of Electromechanical Coupling}
This section will review the basic theory surrounding a superconducting transmon coupled to a diamond defect spin via an intermediary mechanical mode.

We begin by considering a transmon architecture, which consists of a SQUID loop with combined Josephson energy $E_J$ and capacitance $C_J$ in parallel with a shunt capacitor $C_S$. For the sake of constructing only the coupled system, we omit the transmon readout resonator, which typically consists of a quarter wave resonator coupled in parallel to the transmon. The transmon's Josephson and charging energies are $E_J(\phi) = \frac{I_C \Phi_0}{\pi}\cos(\phi_{ext}) = E_J \cos(\phi_{ext})$ and $E_C = \frac{e^2}{2 (C_S + C_J)}$ ($I_C$ is the Josephson junction critical current). Note here that the total charging energy for $\hat{n}$ Cooper pairs will be $4 E_C \hat{n}$, where $\hat{\phi}$ is the conjugate variable of $\hat{n}$. Then the transmon Hamiltonian is given by
\begin{align}
    \hat{H}_{transmon} &= 4 E_C \hat{n} + E_J(\hat{\phi}) \\
    &= 4 E_C \hat{n} + E_J \left(\hat{\phi} + \frac{1}{2}\hat{\phi}^2 + \frac{1}{6}\hat{\phi}^3 + \frac{1}{12}\hat{\phi}^4 + ...\right) \\
    &\approx \left(\sqrt{8 E_J E_C} - E_C\right)\hat{a}^\dag\hat{a} - E_C(\hat{a}^\dag\hat{a}^\dag\hat{a}\hat{a}).
\end{align}

In the last step, we have rewritten in terms of the ladder operators. If we approximate the transmon as a two-level system, then we can simply write $\hat{H}_{transmon}$ as
\begin{equation}
    \hat{H}_{transmon}/\hbar = \frac{\omega_{sc}}{2}\hat{\sigma}^z_{sc}.
\end{equation}

Next, we make note of the Hamiltonian of the electromechanical resonator. Sans coupling, the resonator modes can each be approximated as harmonic oscillators with energy $\hbar\omega_{p,k}$, where $\omega_{p,k}$ is the resonant frequency of the $k$th resonator mode, plus some vacuum energy terms. Ignoring these terms, the Hamiltonian $H_{res}$ is
\begin{equation}
    \hat{H}_{res}/\hbar = \sum_{k}\omega_{p,k}\hat{b}_k^\dag\hat{b}_k.
\end{equation}

Finally, we consider the Hamiltonian of the Group IV electron spin. The full Hamiltonian of Group IV color centers has been discussed at length in \cite{hepp2014electronic}, but for the purposes of this paper we consider the system under an off-axis (transverse and longitudinal) magnetic field (discussed in \cite{neuman2020phononic}). In these conditions, the Group IV Hamiltonian can be written as a sum of the spin-orbit Hamiltonian and a Zeeman perturbation (in the $\{\ket{e_x\uparrow},\ket{e_y\uparrow},\ket{e_x\downarrow},\ket{e_y\downarrow}\}$ basis),
\begin{align}
    \hat{H}_{\textrm{spin}} &= \hat{H}^{\text{SO}} + \hat{H}^{\text{Z}} \\
    &= \begin{bmatrix}
    0&0&-i\lambda_g&0\\
    0&0&0&i\lambda_g\\
    i\lambda_g&0&0&0\\
    0&-i\lambda_g&0&0
    \end{bmatrix} + \begin{bmatrix}
    \gamma_s B_z & \gamma_s B_x & i q \gamma_L B_z & 0\\
    \gamma_s B_x & -\gamma_s B_z & 0 & -i q \gamma_L B_z \\
    -i q \gamma_L B_z & 0 & \gamma_s B_z & \gamma_s B_x \\
    0 & i q \gamma_L B_z & \gamma_s B_x & \gamma_s B_z
    \end{bmatrix} \\
    &= \begin{bmatrix}
    \gamma_s B_z & \gamma_s B_x & -i\lambda & 0\\
    \gamma_s B_x & -\gamma_s B_z & 0 & i\lambda \\
    i\lambda & 0 & \gamma_s B_z & \gamma_s B_x \\
    0 & -i\lambda & \gamma_s B_x & \gamma_s B_z
    \end{bmatrix}. \\
\end{align}
Here, we use $\lambda \equiv \lambda_g - q \gamma_L B_z$ \cite{hepp2014electronic}. Solving the eigensystem of this Hamiltonian gives us the eigenvectors
\begin{multline}
    \ket{\psi_1} = \left(\frac{1}{2\sqrt{\gamma_s^2B_x^2 + (\lambda_{-})\left(\lambda_{-} + \sqrt{\gamma_s^2 B_x^2 + (\lambda_{-})^2}\right)}}\right)\Bigg[\left(-i\left(\lambda_{-} + \sqrt{\gamma_s^2B_x^2 + (\lambda_{-})^2}\right)\right)\ket{e_x\uparrow} + i\ket{e_x\downarrow} \\ -\left(\lambda_{-} + \sqrt{\gamma_s^2B_x^2 + (\lambda_{-})^2}\right)\ket{e_y\uparrow} + \ket{e_y\downarrow}\Bigg],
\end{multline}
\begin{multline}
    \ket{\psi_2} = \left(\frac{1}{2\sqrt{\gamma_s^2B_x^2 + (\lambda_{-})\left(\lambda_{-} + \sqrt{\gamma_s^2 B_x^2 + (\lambda_{-})^2}\right)}}\right)\Bigg[-i\left(\frac{\lambda_{-} - \sqrt{\gamma_s^2B_x^2 + (\lambda_{-})^2}}{\lambda_{-} + \sqrt{\gamma_s^2B_x^2 + (\lambda_{-})^2}}\right)\ket{e_x\uparrow} + i\ket{e_x\downarrow} \\ -\left(\frac{\lambda_{-} - \sqrt{\gamma_s^2B_x^2 + (\lambda_{-})^2}}{\lambda_{-} + \sqrt{\gamma_s^2B_x^2 + (\lambda_{-})^2}}\right)\ket{e_y\uparrow} + \ket{e_y\downarrow}\Bigg],
\end{multline}
\begin{multline}
\ket{\psi_3} = \left(\frac{1}{2\sqrt{\gamma_s^2B_x^2 + (\lambda_{+})\left(\lambda_{+} + \sqrt{\gamma_s^2 B_x^2 + (\lambda_{+})^2}\right)}}\right)\Bigg[-i\left(\frac{\lambda_{+} - \sqrt{\gamma_s^2B_x^2 + (\lambda_{+})^2}}{\lambda_{+} + \sqrt{\gamma_s^2B_x^2 + (\lambda_{+})^2}}\right)\ket{e_x\uparrow} + i\ket{e_x\downarrow} \\ -\left(\frac{\lambda_{+} - \sqrt{\gamma_s^2B_x^2 + (\lambda_{+})^2}}{\lambda_{+} + \sqrt{\gamma_s^2B_x^2 + (\lambda_{+})^2}}\right)\ket{e_y\uparrow} + \ket{e_y\downarrow}\Bigg],
\end{multline}
\begin{multline}
    \ket{\psi_4} = \left(\frac{1}{2\sqrt{\gamma_s^2B_x^2 + (\lambda_{+})\left(\lambda_{+} + \sqrt{\gamma_s^2 B_x^2 + (\lambda_{+})^2}\right)}}\right)\Bigg[\left(-i\left(\lambda_{+} + \sqrt{\gamma_s^2B_x^2 + (\lambda_{+})^2}\right)\right)\ket{e_x\uparrow} - i\ket{e_x\downarrow} \\ +\left(\lambda_{+} + \sqrt{\gamma_s^2B_x^2 + (\lambda_{+})^2}\right)\ket{e_y\uparrow} + \ket{e_y\downarrow}\Bigg].
\end{multline}
These eigenvectors are associated with the eigenvalues
\begin{align}
    \nu_1 &= -\sqrt{\gamma_s^2 B_x^2 + (\lambda_{-})^2},\\
    \nu_2 &= \sqrt{\gamma_s^2 B_x^2 + (\lambda_{-})^2},\\
    \nu_3 &= -\sqrt{\gamma_s^2 B_x^2 + (\lambda_{+})^2},\\
    \nu_4 &= \sqrt{\gamma_s^2 B_x^2 + (\lambda_{+})^2}.
\end{align}

Here, we use $\lambda_{-} = \lambda - \gamma_s B_z$ and $\lambda_{+} = \lambda + \gamma_s B_z$. (Note that, in the limit where $B_x \rightarrow 0$, these eigenvectors and eigenvalues simplify as $\left\{\ket{\psi_1},\ket{\psi_2},\ket{\psi_3},\ket{\psi_4}\right\} \rightarrow \left\{\ket{e_{+}\uparrow},\ket{e_{+}\downarrow},\ket{e_{-}\downarrow},\ket{e_{-}\uparrow}\right\}$ from \cite{hepp2014electronic}).

\begin{figure}
    \centering
    \includegraphics{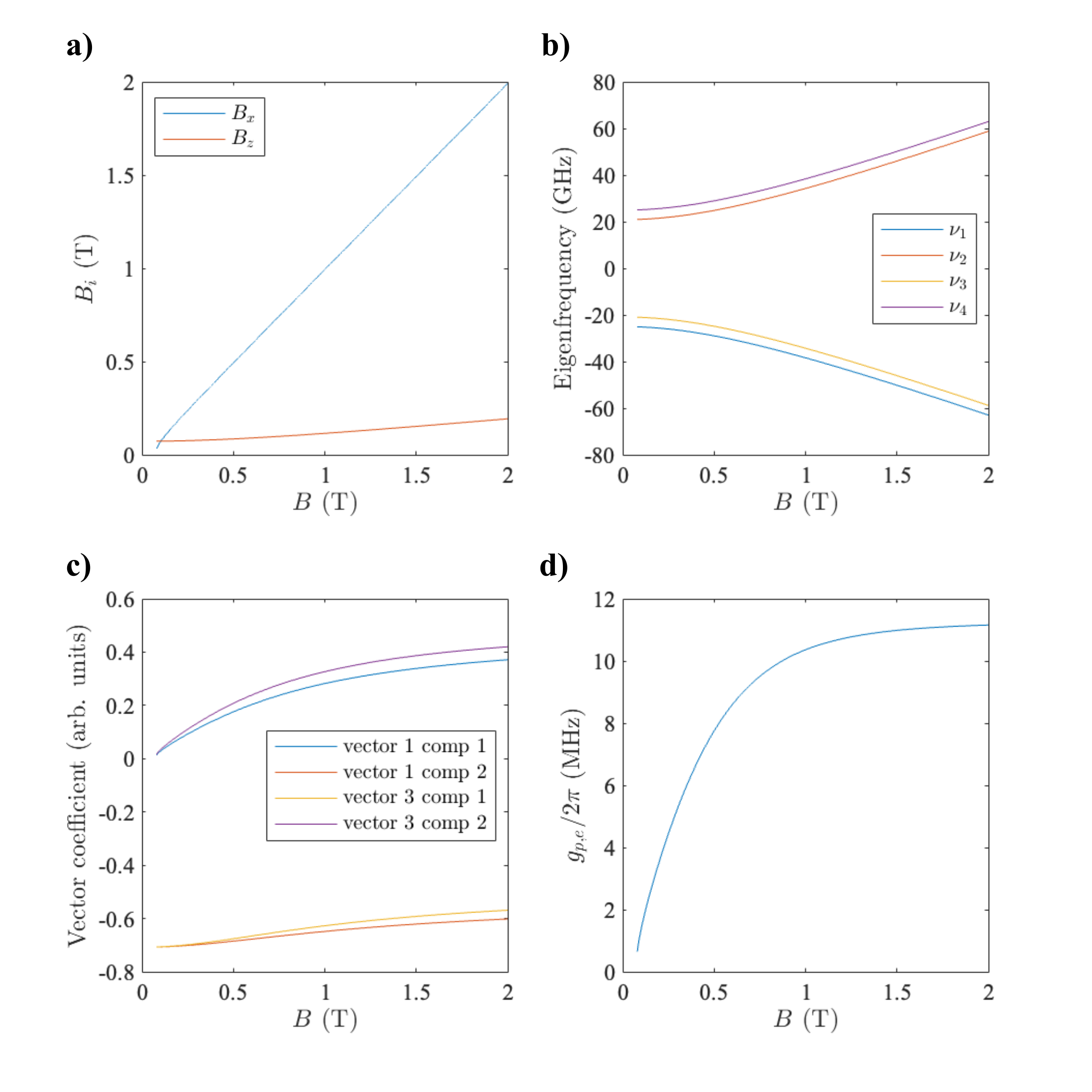}
    \caption{Effect of the maximum applicable magnetic field on various parameters of the system. (a) Evolution of the $B_x$ and $B_z$ required to maintain 4.31 GHz spin splitting as a function of $\abs{B}$. (b) Change in eigenfrequencies as a function of $\abs{B}$, where $\nu_1$ and $\nu_3$ are the eigenfrequencies of $\ket{\psi_1}$ and $\ket{\psi_3}$ are the ground state qubit levels of interest. (c) Change in the components of $\ket{\psi_1}$ and $\ket{\psi_3}$ with $\abs{B}$, indicating greater spin-orbit mixing as the maximum applicable magnetic field increases. (d) projected $g_{p,e}$ vs $\abs{B}$ as determined by Eq.~\eqref{gpe Fermi golden rule}.}
    \label{fig:appendix fig 1 magnetic field effects}
\end{figure}

Finally, the coupling rate $g_{p,e}$ between the lowest lying states $\ket{\psi_1}$ and $\ket{\psi_3}$ can be calculated as
\begin{equation}
    \frac{g_{p,e}}{2\pi} = \abs{\bra{\psi_{3}}M^{-1}H_{strain}M\ket{\psi_1}},
    \label{gpe Fermi golden rule}
\end{equation}
where
\begin{equation}
    H_{strain} = \begin{bmatrix}
    \alpha & 0 & \beta & 0 \\
    0 & \alpha & 0 & \beta \\
    \beta & 0 & -\alpha & 0 \\
    0 & \beta & 0 & -\alpha
    \end{bmatrix}
\end{equation}
and $M$ is the matrix that transforms the eigenvectors $\psi_{i}$ to the strain basis, such that
\begin{equation}
    M \hat{H}_{\textrm{spin}} = M \begin{bmatrix}
    \nu_1 & 0 & 0 & 0\\
    0 & \nu_2 & 0 & 0\\
    0 & 0 & \nu_3 & 0\\
    0 & 0 & 0 & \nu_4
    \end{bmatrix}.
\end{equation}
In SiV$^-$ centers in diamond, $\beta$ is more than ten times smaller than $\alpha$ \cite{meesala2018strain}, so we can simplify $H_{strain}$ to the case where $\beta \rightarrow 0$ and $\alpha \rightarrow \chi_{eff}(\epsilon_{xx} - \epsilon_{yy})$ as discussed in the main text (Equation \eqref{gpe formula}). Then for a known $g_{orb}$ and a maximum magnetic field magnitude $\abs{B}$, we can plot out the required $B_z$ and $B_x$ alongside the projected $g_{p,e}$ (Fig.~\ref{fig:appendix fig 1 magnetic field effects}). We are mostly interested in the regime $0 < \abs{B} \leq 0.2$ T, as this regime lies below the $H_{c_1}$ of Nb. Above this critical field, we would incur additional losses in the coupled system due to the presence of normal currents in the superconducting circuit. As higher $H_{c_1}$ superconductors are explored as SC qubit materials, higher $\abs{B}$ regimes will become accessible to this scheme.


Now, we must consider the coupling between the superconducting circuit and the electron spin to all acoustic modes supported by the piezoelectric resonator. The Hamiltonian of describing this interaction can be written as
\begin{equation}
    \frac{\hat{H}}{\hbar} = \frac{\omega_{sc}}{2} \hat{\sigma}^z_{sc} + \sum_{k} \omega_{p,k} \hat{b}_k^\dag\hat{b}_k + \frac{\omega_e}{2}\hat{\sigma}^z_{e} + \sum_k g_{sc,p;k}\left(\hat{\sigma}^+_{sc} + \hat{\sigma}^-_{sc}\right)\left(\hat{b}_k+\hat{b}_k^\dag\right)+\sum_k g_{p,e;k}\left(\hat{\sigma}^+_{e} + \hat{\sigma}^-_{e}\right)\left(\hat{b}_k+\hat{b}_k^\dag\right).
\end{equation}

where the index $k$ labels each acoustic mode and $\omega_{p,0}$ is the frequency of the resonator mode of interest. We can shift into a interaction picture by applying the transformation $\hat{H}' = \hat{U} \hat{H} \hat{U}^\dag + i \dot{\hat{U}}\hat{U}^\dag$, where $\hat{U} = \exp\left[i\left(\frac{\omega_{sc}}{2}\hat{\sigma}_{sc}^z + \sum_k\omega_{p,k}\hat{b}_k^\dag\hat{b}_k +\frac{\omega_{e}}{2}\hat{\sigma}_{e}^z\right)t\right]$. This transformation gives
\begin{equation}
    \frac{\hat{H}'}{\hbar} = \sum_kg_{sc,p;k}\left(e^{i(\omega_{sc}-\omega_{p,k})t}\hat{\sigma}^+_{sc}\hat{b}_k + e^{-i(\omega_{sc}-\omega_{p,k})t}\hat{\sigma}^-_{sc}\hat{b}_k^\dag\right) + \sum_kg_{p,e;k}\left(e^{i(\omega_{e}-\omega_{p,k})t}\hat{\sigma}^+_{e}\hat{b}_k + e^{-i(\omega_{e}-\omega_{p,k})t}\hat{\sigma}^-_{e}\hat{b}_k^\dag\right), 
    \label{rotating H_int}
\end{equation}

We would like to determine the conditions in which we can neglect all resonator modes except the mode of interest, which we will call $k_0$ with frequency $\omega_{p,k_0}$. Let us first ignore the spin-phonon coupling and focus on the superconducting circuit-phonon coupling. In the interaction Hamiltonian in Eq.~\ref{rotating H_int}, we can see that when $\omega_{sc} = \omega_{p,k_0}$ (the frequency of the acoustic resonator mode of interest), Rabi oscillations will be induced between the two modes. We would also, however, like to consider the oscillations induced between the superconducting circuit and the other resonator modes. Let us select a different transformation $\hat{H}'_2 = \hat{U}_2 \hat{H}\hat{U}_2^\dag + i\dot{\hat{U}}_2\hat{U}_2^\dag$, where $\hat{U}_2 = \exp\left[i\left(\frac{\omega_{sc}}{2}\hat{\sigma}_{sc}^z + \sum_k\left(\omega_{p,k} + \Delta_{p,k}\right)\hat{b}_k^\dag\hat{b}_k\right)t\right]$, where $\Delta_{p,k}\equiv \omega_{sc} - \omega_{p,k}$, and ignore the electron spin-related terms. The resulting interaction Hamiltonian is
\begin{equation}
    \frac{\hat{H}'_2}{\hbar} = -\sum_{k}\Delta_{p,k}\hat{b}_k^\dag\hat{b}_k + \sum_{k} g_{sc,p;k}\left(\hat{\sigma}^+_{sc}\hat{b}_k + \hat{\sigma}^-_{sc}\hat{b}_k^\dag\right).
\end{equation}

The Heisenberg equations of motion for $\hat{\sigma}_{sc}$ and $\hat{b}_{k}$ are
\begin{align}
    \dot{\hat{\sigma}}^-_{sc} &= -\frac{i}{\hbar}\left[\hat{H}_2',\hat{\sigma}^-_{sc}\right]\\
    &= -\frac{\kappa_{sc}}{2}\hat{\sigma}^-_{sc} - i g_{sc,p;k_0}\hat{b}_{k_0} - i\sum_{k\neq k_0} g_{sc,p;k}\hat{b}_k,
    \\
    \dot{\hat{b}}_{k} &= -\frac{i}{\hbar}\left[\hat{H}_2',\hat{b}_{k}\right] \\
    &= \left(-i\Delta_{p,k} - \frac{\kappa_{p,k}}{2} \right)\hat{b}_{k} + i g_{sc,p;k}\hat{\sigma}^-_{sc}
\end{align}
where $g_{sc,p;k_0}$ is the desired acoustic mode's electromechanical coupling. 
In matrix form, this becomes
\begin{equation}
    \dot{\begin{bmatrix}
    \hat{\sigma_{sc}}\\
    \hat{b_{1}}\\
    \hat{b_{2}}\\
    \vdots\\
    \hat{b_{N}}\\
    \end{bmatrix}} = \begin{bmatrix}
    -\frac{\kappa_{sc}}{2} & -i g_{sc,p;1} & -i g_{sc,p;2} & \ldots & -i g_{sc,p;N}\\
    i g_{sc,p;1} & (-i\Delta_{p,1}-\frac{\kappa_{p,1}}{2}) & 0 & \ddots & 0 \\
    i g_{sc,p;2} & 0 & (-i\Delta_{p,1}-\frac{\kappa_{p,1}}{2}) & \ddots & 0\\
    \vdots & \ddots & \ddots & \ddots & \vdots \\
    i g_{sc,p;N} & \ldots & \ldots & \ldots & (-i\Delta_{p,N}-\frac{\kappa_{p,N}}{2})
    \end{bmatrix}\begin{bmatrix}
    \hat{\sigma_{sc}}\\
    \hat{b_{1}}\\
    \hat{b_{2}}\\
    \vdots\\
    \hat{b_{N}}\\
    \end{bmatrix}.
\end{equation}

This is equivalent to inducing Rabi oscillations of various frequencies and suppressions between the SC qubit and acoustic modes. The probability amplitude of population transfer to each acoustic mode from an excited SC state becomes
\begin{equation}
    \langle \sigma_{sc,k} \rangle = \frac{4 (g_{sc,p;k})^2}{4(g_{sc,p;k}^2)+\abs{\Delta_{p,k} + i\left(\frac{\kappa_{sc}+\kappa_{p,k}}{2}\right)}^2}\sin^2\left(\frac{\sqrt{4 \left(g_{sc,p;k}\right)^2 + \abs{\Delta_{p,k} + i\left(\frac{\kappa_{sc}+\kappa_{p,k}}{2}\right)}^2}}{2} t\right).
    \label{Rabi pop}
\end{equation}

This gives us a SC qubit probability of being in the excited state as a function of time is then 
\begin{equation}
    \hat{\sigma}_{sc} = \sum_{k}\langle \sigma_{sc,k}\rangle = \sum_{k}\frac{4(g_{sc,p;k})^2}{4 \left(g_{sc,p;k}\right)^2 + \left(\Delta_{p,k} + i\left(\frac{\kappa_{sc}+\kappa_{p,k}}{2}\right)\right)^2}\sin^2\left(\frac{\sqrt{4 \left(g_{sc,p;k}\right)^2 + \abs{\Delta_{p,k} + i\left(\frac{\kappa_{sc}+\kappa_{p,k}}{2}\right)}^2}}{2} t\right).
    \label{Rabi pop total}
\end{equation}

The sum over all $\langle \sigma_{sc,k} \rangle$ with $k \neq k_0$ is a worst-case bound on the probability amplitude that could escape the computational basis into undesired acoustic modes, limiting state fidelity. If the ratio of $\langle \sigma_{sc,k_0}\rangle/\sum_{k\neq k_0}\langle \sigma_{sc,k}\rangle \gg 1$, then we can effectively treat our system as having only one acoustic mode coupled to a SC qubit. The same physics governs the spin-phonon dynamics, replacing the appropriate couplings in equation \eqref{Rabi pop} and \eqref{Rabi pop total}. 

\section{Details of Numerical Simulations}

Simulations were completed using the finite element method (FEM) in COMSOL Multiphysics, utilizing the Electrostatics and Structural Mechanics modules. Stationary simulations were conducted to determine the electrostatic field applied to the piezoelectric transducer from a microwave source, and eigenfrequency simulations were conducted to determine the transducer's resonant acoustic modes. Finally, $g_{sc,p}$ was calculated for each mode by determining the overlap between the piezoelectrically induced field and electrostatic field (see Eq.
\begin{figure}
    \centering
    \includegraphics[width=\textwidth]{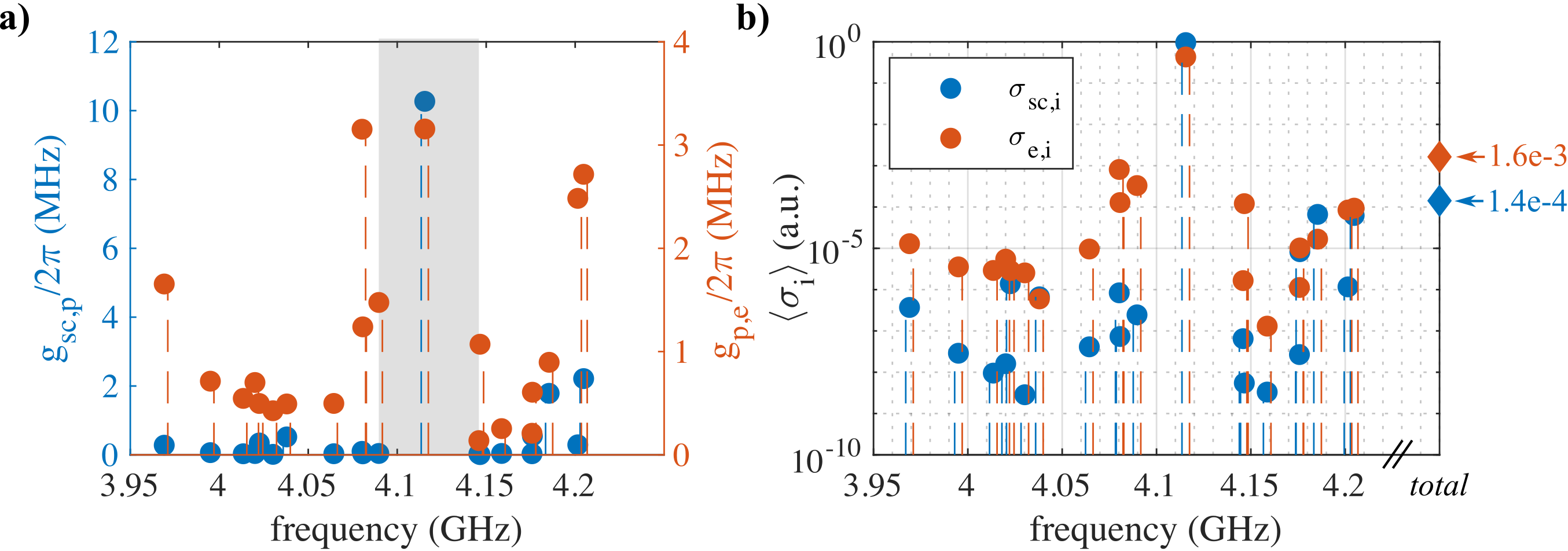}
    \caption{Electromechanical and spin-mechanical couplings and population transfer to each acoustic mode. (a) demonstrates a $\sim56$ MHz frequency window (grey shaded region) in which our mode of interest ($\sim4.115$ GHz) lies. The couplings $g_{sc,p_i}$ and $g_{p,e_i}$ are plotted for each mode, assuming a shunt capacitance $C_S \sim 130$ fF and a magnetic field of $0.18$ T. (b) displays the Rabi population transfer probability from the superconducting circuit and electron spin to each acoustic mode (see Eq.~\eqref{Rabi pop}), showing a combined mode suppression (diamond markers) of at least three orders of magnitude}
    \label{fig:couplings and pops}
\end{figure}

\begin{figure}
    \centering
    \includegraphics[width=\textwidth]{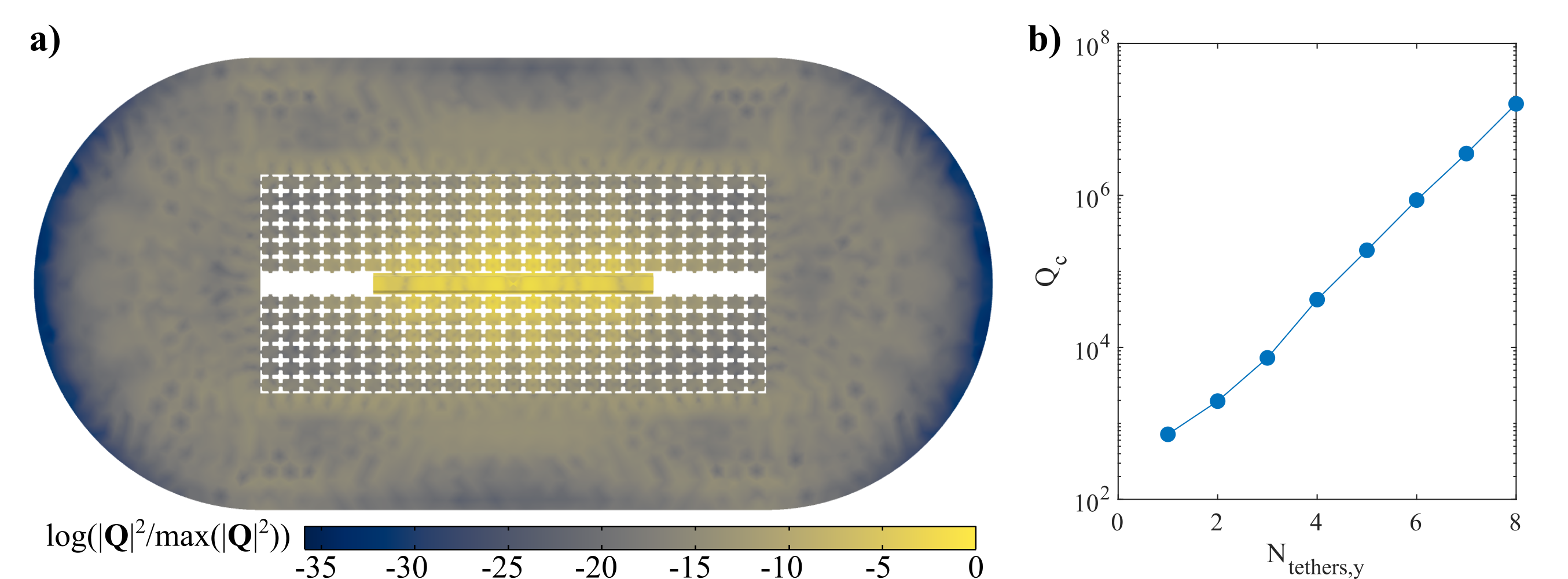}
    \caption{FEM simulation of the piezoelectric transducer with phononic tethers and surrounding bulk treated as perfectly matched layers (PML) to simulate clamping quality factor $Q_c$. (a) Diagram of the setup with free variable $N_{tethers,y}$ identified and (b) plot of $Q_c$ vs $N_{tethers,y}$ for the 4.11 GHz resonator mode of interest.}
    \label{fig: Q vs tethers}
\end{figure}

The coupling parameters $g_{sc,p}$ and $g_{p,e}$ were then calculated using the combination of these two simulations (see Eqs. ~\eqref{gscp formula} and ~\eqref{gpe formula}). The parameter set with the best mode isolation (see Fig.~\ref{fig:couplings and pops} featured $\lambda = 1370$ nm and $w_{res}=465$ nm. This device was then tethered using the phononic tethers shown in Fig.~\ref{fig 2: diagram} and the number of tethers were varied to calculate mechanical quality factor as a function of number of tether periods, shown in Fig.~\ref{fig: Q vs tethers}(b).

\section{Analysis of Spin Register System}

\begin{figure}
    \centering
    \includegraphics[width=\linewidth]{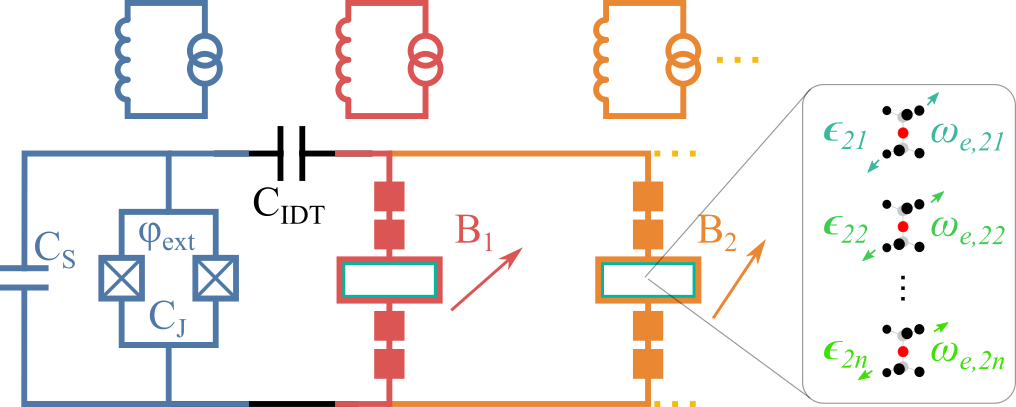}
    \caption{Scaling the schematic to a quantum memory register. By implanting $n$ emitters in each of $m$ detuned mechanical resonators in parallel with the supercondcuting qubit of interest, one can create an efficient interface to an $m \times n$ optically addressable ancilla register.}
    \label{fig 5: outlook}
\end{figure}

In Figure \ref{fig 5: outlook}, we present a roadmap to scaling this architecture to form a memory register for superconducting circuits. Since the shunt capacitance far exceeds the capacitance of a single IDT, additional electromechanical resonators in parallel to a single transmon qubit do not significantly change the coupling rates to each resonator. Individual control over each resonator can be obtained with (i.e. cryo-MEMS) electrical switching of contacts to each resonator \cite{brown2008cryogenic}. If this is not possible, controls can still be obtained in the frequency domain if each resonator frequency is sufficiently detuned from all others and within the tunability range of the transmon. This gives two constraints on the number of parallel resonators we can add: the maximum number of resonators before $g_{sc,p}$ for each resonator drops below a desired value, and the maximum number of resonators before the frequency spectrum becomes overcrowded.

From electrostatic simulations in COMSOL, $C_S \approx 70 C_{IDT}$, allowing us to add around 10 resonators in parallel without decreasing the coupling to each resonator by more than 15\%. Additionally, each resonator can house several quantum emitters, which themselves will be operating at different frequencies $\omega_{e,ij}$ due to differing magnetic field and strain environments creating a unique Zeeman effect for each color center. Assuming one implants $n$ emitters in each resonator, this creates an easily accessible $m \times n$ register of ancillas for a single transmon.

We would like to evaluate overcrowding of the frequency spectrum in this picture. In an ideal case, when we tune the superconducting circuit on resonance with a mechanical mode $\omega_m$, we would like the circuit to be approximately coupled \textit{only} to that acoustic mode. This is the same condition as we presented in Appendix A to assume that we can simplify the dynamics of the SC-phonon-spin system to that of coupling via a single acoustic mode. Thus, when the condition
for every mode $j$, then we can suppose that we individually couple to one piezoelectric resonator out of a number of resonators (see Fig.~\ref{fig 5: outlook}). Similarly, we would like to determine the condition where we can assume each piezoelectric resonator can individually couple to a single spin. This complicates the second stage of the system in Appendix A. Assuming that the conditions in Appendix A already holds for each of $m$ resonators coupled to the SC qubit, the full Hamiltonian describing the $m$ resonator, $m \times n$ spin system is
\begin{equation}
    H_{\sum} =\frac{\omega_{sc}}{2}\hat{\sigma}^{z}_{sc}  + \sum_{i = 1}^m\sum_{j=1}^n\Bigg[\omega_{p,i} \hat{a}_{p,i}^\dag \hat{a}_{p,i} + \frac{\omega_{e_{ij}}}{2}\hat{\sigma}^{z}_{e_{ij}}
    + g_{sc,p_i}\left( \hat{\sigma}^{+}_{sc} \hat{a}_{p,i} + \hat{\sigma}^{-}_{sc}\hat{a}_{p,i}^\dag\right)
    + g_{p,e_{ij}}\left(\hat{\sigma}^{+}_{e_{ij}}\hat{a}_{p,i} + \hat{\sigma}^{-}_{e_{ij}}\hat{a}_{p,i}^\dag\right)\Bigg].
\end{equation}

Following exactly from Eqs.~\ref{Rabi pop} and \ref{Rabi pop total} in Appendix A, the required condition for assuming electromechanical coupling to just the $k_0$th of $m$ resonators is that
\begin{equation}
    \sum_{k\neq k_0}^m \langle \sigma_{sc,k} \rangle = \sum_{k \neq k_0}^m\frac{4 (g_{sc,p;k})^2}{4(g_{sc,p;k}^2)+\abs{\Delta_{p,k} + i\left(\frac{\kappa_{sc}+\kappa_{p,k}}{2}\right)}^2}\sin^2\left(\frac{\sqrt{4 \left(g_{sc,p;k}\right)^2 + \abs{\Delta_{p,k} + i\left(\frac{\kappa_{sc}+\kappa_{p,k}}{2}\right)}^2}}{2} t\right) \ll \langle \sigma_{sc,k_0}\rangle.
\end{equation}
Similarly, after swapping population into one of the resonator modes, the condition for assuming spin-mechanical coupling to just the $j_0$th of $n$ electron spins is that
\begin{equation}
    \sum_{j\neq j_0}^n \langle \sigma_{e,j} \rangle = \sum_{j \neq j_0}^n\frac{4 (g_{p,e;j})^2}{4(g_{p,e;j}^2)+\abs{\Delta_{p,j} + i\left(\frac{\kappa_{e}+\kappa_{p,j}}{2}\right)}^2}\sin^2\left(\frac{\sqrt{4 \left(g_{p,e;j}\right)^2 + \abs{\Delta_{p,j} + i\left(\frac{\kappa_{e}+\kappa_{p,j}}{2}\right)}^2}}{2} t\right) \ll \langle \sigma_{e,j_0}\rangle.
\end{equation}



We can see from the spin-phonon coupling points in Fig.~\ref{fig:couplings and pops} that frequency crowding can begin to promote Rabi oscillations with populations on the order of $10^-3$ of the desired mode when within a 100 MHz frequency window. So parallelization of spins in one resonator would require changing the local magnetic field for each resonator and intelligent spacing of the emitters to promote a wide distribution of resonant frequencies, or sacrificing state tansfer fidelity to a single spin by overcrowding the simulated frequency window of operation. This is not as much of a problem given the order-of-magnitude superior mode suppression on the electromechanical side of the system. Thus, we can comfortably parallelize around $10$ piezoelectric resonators to a single SC qubit and 1-3 emitters per resonator. When accounting for the surrounding $^{13}C$ nuclear spins, we envision that this scaling method can provide a SC qubit with a 10+ nuclear spin memory register.

\end{document}